\documentstyle[prl,aps,epsf]{revtex}
\begin{document}

\advance\textheight by 0.5in
\advance\topmargin by -0.25in
\draft

\twocolumn[\hsize\textwidth\columnwidth\hsize\csname@twocolumnfalse%
\endcsname

\preprint{cond-mat/9902159}       

\title{\hfill {\small cond-mat/9902159}  \\ 
\vspace{10pt}
``X-Ray Edge'' Singularities in Nanotubes and Quantum Wires with Multiple Subbands}
\author{Leon Balents}
\address{Bell Labs, Lucent Technologies, Room 1D-368, 700 Mountain
  Ave, Murray Hill, NJ 07974} 
\date{\today}
\maketitle


\begin{abstract}
Band theory predicts a $(\epsilon-\Delta)^{-1/2}$ van
Hove singularity in the tunneling density of states at the minimum
energy ($\Delta$) of an unoccupied subband in a one-dimensional
quantum wire.  With interactions, an
orthogonality catastrophe analogous to the x-ray edge effect for core
levels in a metal strongly reduces this singularity to the form
$(\epsilon-\Delta)^{\beta-1/2}$, with $\beta \approx 0.3$ for typical
carbon nanotubes.  Despite the anomalous tunneling characteristic,
good quasiparticles corresponding to the unoccupied subband states do
exist. 
\end{abstract}
\vskip -0.5 truein
]

\section{Introduction}

It is well known that one-dimensional (1d) metals are proundly
influenced by interactions.  The generic behavior for a 1d system with
a single conduction band is that of a Luttinger liquid, in which the
quasiparticle excitations of the non-interacting system are converted
into spin and charge collective modes.  These collective modes are
largely orthogonal to the bare quasiparticle states close to the Fermi
surface, hence leading to a dramatic power-law suppression of the
tunneling Density of States (DOS) at the Fermi energy.  A similar vanishing
occurs in the momentum-dependent single-particle spectral function,
signaling complete breakdown of the quasiparticle picture.  Indeed,
tunneling experiments evidence the novel nature of the Luttinger
liquid much more strongly than do other measures such as four-probe
resistance and optical conductivity, which are essentially probes of
collective modes.  There now appears to be increasing evidence for
Luttinger liquid behavior in carbon nanotubes,\cite{Bockrath99}\ which
are perhaps the most ideal experimental quantum wires to
date.\cite{Wires,Yacoby96}\ 


In this paper, we address the behavior of the tunneling DOS {\sl far
  above} the Fermi energy.  As the voltage bias between a tunneling
tip and the quantum wire is increased, it becomes possible to add an
electron not only to the conduction band(s), but to higher energy {\sl
  unoccupied} subbands.  Such multiple subband structure exists both
in carbon nanotubes and semiconductor quantum wires.  In the former
case, the unoccupied subbands are essentially states of different
angular momentum around the graphitic cylinder.  For semiconductor
systems, the higher subbands arise from different standing-wave modes
transverse to the wire's propagation direction in the confinement
region.  In a non-interacting model, the density of states would
exhibit van Hove singularities at the subband edges.  In one
dimension, these singularities are divergent, giving a contribution
$\rho_0(\epsilon) \sim \sqrt{m} (\epsilon-\Delta)^{-1/2}
\Theta(\epsilon-\Delta)$ for energies just above the subband edge at
$\epsilon=\Delta$ ($m$ is the subband effective mass).  An asymmetric
peak structure has indeed been observed in STM measurements of
individual nanotubes on gold surfaces.\cite{Wildoer98,Odom98}\ 

How do interactions affect these van Hove singularities?  A
simplified, though unphysical model in which the mass of the higher
subband is taken to be infinite provides considerable insight.  In
this limit the higher energy subbands can be replaced by discrete,
localized levels.  The ``x-ray edge'' problem of a localized level
interacting with a conduction sea was solved by Nozieres and de
Dominicis\cite{Nozieres69}, and is one of the first demonstrations of an orthogonality
catastrophe.  Physically, the core hole is ``dressed'' through
interactions with conduction electrons, which see the hole as a
scattering center.  A bare or undressed hole is then orthogonal to its
dressed counterpart, since an infinite number of conduction electrons
are available to scatter off of it.  This leads to a broadening and
reduction of the tunneling density of states from a sharp
delta-function to a power law singularity.  While this effect is
superficially similar to the suppression of spectral weight in the
Luttinger liquid, it is in fact quite distinct.  It is due not to the
absence of a well-defined core hole excitation, but due to its mixing
with the conduction sea.  The distinction is emphasized by the fact
that x-ray edge singularities are present in any dimension, not just
in 1d.

Suppose now that the mass of the higher subband is taken finite.
Since for the tunneling density of states only a single particle is
being added to the metal, one is faced with understanding the behavior
of a heavy particle in a Luttinger liquid.  This problem has been
investigated by a number of authors in a different
context.\cite{Moustakas95,Neto96}\ These authors were primarily
concerned with the {\sl mobility} of the heavy particle in response to
an external electric field at finite temperature.  For
the tunneling DOS, one is interested in a rather
different property -- essentially, the overlap between the two ground
states in the presence and absence of the heavy particle. This overlap may be
thought of as a boundary condition changing operator\cite{Affleck94}, and is
completely outside the Hilbert space of the problem in which the heavy
particle is always present.

In this paper, we demonstrate that x-ray edge effects persist even for
this finite mass case.  These reduce the naive van Hove singularities
in the tunneling DOS by a power-law amount.  Like the
singularities in the original x-ray edge problem (but {\sl unlike} the
the Luttinger singularities at low bias), this modification does not,
however, signal the destruction of sharp quasiparticle excitations in
the higher subbands.  We argue that a necessary and sufficient
condition for the presence of such finite-energy singularities is a
conserved quantum number distinguishing the states of the higher
subband from the conduction states.  In the case of the carbon
nanotube, this is an angular momentum quanta.  For a semiconductor
wire, such a good quantum number exists in the ideal case of a
symmetric confining potential, in which case the second
subband has odd parity with respect to reflection, while the
lowest subband has even parity.  If such a distinguishing quantum
number is absent, we expect the van Hove peak to be rounded and 
rendered completely nonsingular.  

The case when only {\sl forward scattering} interactions are present
between the higher subband particles and the conduction sea is
asymptotically exactly soluble by bosonization methods, as we outline
here.  As discussed in Ref.~\onlinecite{KBF97}, this forward
scattering model is in fact an excellent approximation for
single-walled carbon nanotubes with diameters $D \gtrsim 1 nm$.
Within this model, we predict a reduced density of states singularity
\begin{equation}
  \rho(\epsilon) \sim \rho_0 \left( \Delta \over {\epsilon-\Delta}
  \right)^{{1 \over 2} - \beta} \Theta(\epsilon-\Delta),
  \label{dos}
\end{equation}
where $\Theta(x)$ is the heavyside step function, and the
orthogonality exponent $\beta \approx 0.3$ for typical metallic
nanotubes.  The form of Eq.~\ref{dos}\ is expected to apply to the
second subband in semiconductor quantum wires as well (see the
concluding remarks for a discussion of experiments).

\section{Forward Scattering Charge Model}

In this section we present a simple Forward Scattering Charge Model (FSCM)
describing only forward scattering interactions, i.e. those processes
involving small momentum exchange, in the total charge channel.  We describe the conduction
electrons by a Luttinger model, valid near the Fermi energy, which we
take to be $E=0$,
\begin{equation}
  H_0^c = \sum_{i\alpha} \int\! dx\, v_F \left[ \psi_{Ri\alpha}^\dagger
    i\partial_x \psi_{Ri\alpha}^{\vphantom\dagger} - \psi_{Li\alpha}^\dagger
    i\partial_x \psi_{Li\alpha}^{\vphantom\dagger} \right].
  \label{eq:luttinger}
\end{equation}
Here $v_F$ is the Fermi velocity, and $\alpha = \uparrow,\downarrow$
labels the electron spin.  We have also included an additional
``flavor'' index $i=1\ldots N_f$ to allow for extra degenerate bands
(with the same Fermi velocity) at the Fermi energy.  For metallic
carbon nanotubes, $N_f = 2$ and $i$ should be interpreted as a
sublattice index.  No such special degeneracy is present for a
semiconductor quantum wire, so $N_f=1$ in this case.
Eq.~\ref{eq:luttinger}\ can be rewritten using bosonization.  We
follow the conventions of Ref.~\onlinecite{KBF97}.  One has
$\psi_{R/L;i\alpha} \sim e^{i(\phi_{i\alpha} \pm \theta_{i\alpha})}$,
where the dual fields satisfy $[\phi_{i\alpha}(x), \theta_{j\beta}(y)]
= -i\pi \delta_{ij}\delta_{\alpha\beta} \Theta(x-y)$ ($\Theta(x)$ is a
heavyside step function).  Then $H_0^c = \sum_{i,\alpha} \int\! dx\, {\cal
  H}_0^c(\theta_{i\alpha},\phi_{i\alpha})$, with
\begin{equation}
  {\cal H}_0^c(\theta,\phi) ={v_F \over {2\pi}} [(\partial_x
  \theta)^2 + (\partial_x \phi)^2].
  \label{Hspin}
\end{equation}
The slowly varying electronic density in a given channel is given by
$\rho_{i\alpha} \equiv \psi_{Ri\alpha}^\dagger
\psi_{Ri\alpha}^{\vphantom\dag} + \psi_{Li\alpha}^\dagger
\psi_{Li\alpha}^{\vphantom\dagger} = \partial_x\theta_{i\alpha}/\pi$.
Physically, $\theta$ can be understood as a displacement or phonon
field, while $\phi$ carries the phase of the quantum wavefunction.

It is simplest to work in a rotated basis of collective modes.  For
$N_f=1$, define $\theta_{\rho/\sigma} = (\theta_\uparrow \pm
\theta_\downarrow)/\sqrt{2}$.  For $N_f=2$,  let $\theta_{i,\rho/\sigma} = (\theta_{i\uparrow} \pm
\theta_{i\downarrow})/\sqrt{2}$ and $\theta_{\mu\pm} =(\theta_{1\mu}\pm 
\theta_{2\mu})/\sqrt{2}$, with $\mu=\rho,\sigma$.  With similar
definitions for the $\phi$ fields, canonical commutators are
preserved, and $H_0^c = \sum_a \int_x {\cal H}_0^c(\theta_a,\phi_a)$,
where $a$ is summed over the $2N_f$ rotated boson fields.

Because we are interested only in energies near the putative van Hove
singularity, the unoccupied 1d subband can be described by a
non-relativistic electron operator $d,d^\dagger$:
\begin{equation}
  H_0^d = \int \! dx\, d_\alpha^\dagger \left[ -{1 \over
      {2m}}\partial_x^2 + \Delta\right] d^{\vphantom\dagger}_\alpha.
  \label{eq:heavy}
\end{equation} 
Here $\Delta$ is the gap to the first subband and $m$ is an effective
mass.  The electron field satisfies
$\{d_\alpha^{\vphantom\dagger}(x),d_\beta^\dagger(x')\} =
\delta_{\alpha\beta}\delta(x-x')$.  In the case of a carbon nanotube,
there are actually multiple degenerate subbands at energy $\Delta$.
This degeneracy is unimportant within the FSCM, as the tunneling DOS
involves only states with a single excited electron We therefore
neglect this additional degeneracy. 

The interactions in the FSCM are written as a single term coupling
only the total charge density, 
\begin{equation}
  H_{\rm int} = {1 \over 2}\int\! dx dx' \rho_{\rm tot}(x) V(x-x')
  \rho_{\rm tot}(x'),
  \label{eq:nonlocal}
\end{equation}
where
\begin{equation}
  \rho_{\rm tot} = -e(d^\dagger d + \sum_{i\alpha} \rho_{i\alpha}) =
  -e(d^\dagger d + {\sqrt{2N_f} \over \pi} \partial_x\theta_{\rho}).
\end{equation}
Here and in what follows we abbreviate $\theta_{\rho+} = \theta_\rho$
for the $N_f=2$ case.  A phenomenological form for the potential is
sufficient for our purposes.  We take $V(x) =
\exp(-|x|/\xi)/\sqrt{x^2+W^2}$, modeling the smoothing of the
interaction on the scale of the wire width by $W$ and including a
screening length $\xi$, determined, e.g. by the distance to an
external gate (any dielectric constant can be included by rescaling
$e^2 \rightarrow e^2/\epsilon$.

While it is possible to proceed directly with the non-local form in
Eq.~\ref{eq:nonlocal}, near to the van Hove singularity (within an
energy of order $v_F/W$, up to a weak logarithmic factor) it is
sufficient to make the local approximation $V(x) \rightarrow
\delta(x)\int\! dx' V(x')$.  This gives $H_{\rm int} = \int_x {\cal
  H}_{\rm int}$, with
\begin{equation}
  {\cal H}_{\rm int} = e^2 \ln(\xi/W) \left({\sqrt{2N_f} \over \pi}
    \partial_x\theta_\rho + d^\dagger d\right)^2.
\end{equation}

\section{Solution of FSCM}

To determine the effects of the interaction, it is convenient to
employ a path integral formulation.  Quantum mechanical expectation
values are evaluated as functional integrals over classical fields in
imaginary time with respect to a measure $\exp(-\int \! dx d\tau {\cal 
  L})$, where ${\cal L}$ is a Lagrange density.  Standard techniques
give 
\begin{equation}
  {\cal L} = {i \over \pi} \partial_x\theta_\rho
  \partial_\tau\phi_\rho + d^\dagger\partial_\tau d^{\vphantom\dagger} 
  + {\cal H},
  \label{eq:lagrange1}
\end{equation}
with
\begin{eqnarray}
  {\cal H} & = & {\cal H}_0^c + {\cal H}_0^d + {\cal H}_{\rm int}
  \nonumber \\
  & = & {v_\rho \over {2\pi g}}\left[\partial_x\theta_\rho + \gamma d^\dagger d^{\vphantom\dagger}
  \right]^2 + {{gv_\rho} \over {2\pi}}(\partial_x\phi)^2 + {\cal
    H}_0^d.
  \label{eq:bosonmodel}
\end{eqnarray}
Here we have defined the plasmon velocity $v_\rho =
\sqrt{v_F(v_F+(4N_f e^2/\pi\hbar)\ln (\xi/W))}$, Luttinger parameter
(``conductance'') $g=v_F/v_\rho$, and $\gamma = {\pi \over
  \sqrt{2N_f}}(1-g^2)$.

It may be tempting to proceed by perturbation theory in $\gamma$.
Indeed, for properties of the conduction electrons at energies small
compared to $\Delta$, this is a perfectly sensible procedure: as no
heavy particles are present, the properties of the conduction sea are
completely unaffected.  However, the same is not true for the
tunneling density of states.  This is obtained from the heavy particle
Green's function, $G(x,\tau) \equiv \langle
d^{\vphantom\dagger}(x,\tau) d^\dagger(0,0)\rangle$, via
$\rho(\epsilon) = \pi^{-1} {\rm Im} \int \! dk/2\pi G(k,i\omega
\rightarrow \omega + i0^+ )$.  The first perturbative correction to
$G$, obtained from the self-energy diagram in Fig.~1a, is
logarithmically divergent.  Although we will not proceed along this
route, this logarithmic divergence can be controlled to leading order
using a renormalization group procedure which treats in a
self-consistent fashion both this self-energy correction and the
additional vertex renormalization given by the diagram in Fig.~1b.
The results of this calculation\cite{unpublished}\ are confirmed by a
non-perturbative analysis, to which we now turn.
\begin{figure}[htb]
\epsfxsize=3.3in\epsffile{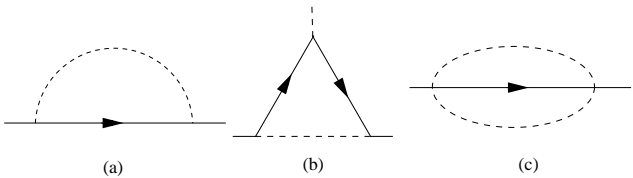}\vspace{0.1in}
\caption{Diagrams in perturbation theory.  A direct attack
  requires a renormalization group resummation of diagrams (a) and
  (b), where the solid line is the heavy-particle propagator, the
  dashed line is the $\theta_\rho$ propagator, and the vertices each
  carry a momentum factor.  After the
  transformation of Eqs.~\ref{eq:densityshift}-\ref{eq:string}, the
  residual interactions in Eq.~\ref{eq:residual}\ generate only
  irrelevant corrections via diagrams (a) and (c), where now the
  dashed lines represent $\phi_\rho$ propagators and the vertices
  carry {\sl two} momentum factors each.}
\end{figure}
The model in Eq.~\ref{eq:bosonmodel}\ is solved by a canonical
transformation, or change of variables in the path integral: 
\begin{eqnarray}
  \theta_\rho(x) & = & \tilde\theta_\rho(x) - \gamma  \int^x_{-\infty} \! dx'
  \, d^\dagger(x') d^{\vphantom\dagger}(x'), \label{eq:densityshift}\\
  d(x) & = & e^{i\gamma\phi_\rho(x)/\pi} \tilde{d}(x). \label{eq:string}
\end{eqnarray}
Eqs.~\ref{eq:densityshift}-\ref{eq:string}\ embody the physical
process in which the conduction sea {\sl adiabatically adjusts to the
  heavy particle}.   In particular, Eq.~\ref{eq:densityshift}\
represents the depletion of the conduction electron density near the
heavy particle due to Coulomb repulsion.  Eq.~\ref{eq:string}\
represents phase shifts of these conduction electrons when the heavy
particle is introduced.  Formally, the exponential of the dual
($\phi$) field in Eq.~\ref{eq:string}\ is a Jordan-Wigner ``string''
operator which has been attached to the heavy particle.

In the new variables, the hamiltonian density becomes
\begin{equation}
  {\cal H}  =  {v_\rho \over {2\pi g}}(\partial_x\tilde\theta_\rho)^2 +
  {{gv_\rho} \over {2\pi}}(\partial_x\phi)^2 + {\cal
    H}_0^d[\tilde{d},\tilde{d}^\dagger] + \tilde{\cal H}_{\rm int},
  \label{eq:hamtilde}
\end{equation}  
with a different residual interaction term
\begin{equation}
  \tilde{\cal H}_{\rm int} = \left[ {\gamma^2 \over {2m\pi^2}}
    (\partial_x\phi_\rho)^2 - i {\gamma \over {2m\pi}}
    \partial_x^2\phi_\rho \right] \tilde{d}^\dagger
  \tilde{d}^{\vphantom\dagger}.
  \label{eq:residual}
\end{equation}
It might appear that the transformations have actually worsened the
problem, as the interaction in Eq.~\ref{eq:residual}\ naively appears
more complicated than the original in Eq.~\ref{eq:bosonmodel}.
However, a closer inspection shows that the new couplings in
Eq.~\ref{eq:residual}\ are dimensionally weaker by one inverse power
of length than the original forms.  As the original interaction was
marginal, the terms in Eq.~\ref{eq:residual}\ are in fact {\sl
  irrelevant} in the renormalization group sense.  This can be
verified by an explicit calculation of their effects on the
$\tilde{d}$ Green's function.  Apart from a constant renormalization
of the subband gap, the leading order diagrams (Figs.~1a,1c) give
self-energy contributions proportional to $(\omega-\Delta)^3$.  

The irrelevance of the couplings in Eq.~\ref{eq:residual}\ indicates
that at long times and distances, the transformed fermion and boson
correlation functions asymptotically factorize.  Thus at energies
close to the threshold energy $\Delta$, the $\tilde{d}^\dagger$
operator creates good quasiparticles which propagate independently of
the conduction sea.  The tunneling DOS, however, involves the addition
of a {\sl bare} electron created by the $d^\dagger$ operator.
Factorization implies
\begin{equation}
  G(x,\tau) = G^0(x,\tau)/\left|\Lambda\sqrt{x^2+v_\rho^2\tau^2}\right|^\beta,
\end{equation}
where $G^0$ is the non-interacting Green's function describing free
propagation in the unoccupied subband, $\beta = \gamma^2/2\pi^2g =
(1-g^2)^2/(4N_f g)$, and $\Lambda$ is a momentum cutoff ($O(k_F)$).
When Fourier transformed, the space-time product above becomes a
convolution, which physically represents the emission of plasmon waves 
by the added electron.  Analytic continuation to real frequency
gives the modified van Hove singularity in Eq.~\ref{dos}.

\section{Discussion}

The preceding analysis demonstrates the persistance of a well-defined
edge in the tunneling spectrum in the presence of forward scattering
charge interactions with the conduction electrons.  The existence of
such a finite energy singularity hinges on the inability of the heavy
particle to truly decay or mix with the many other (conduction
subband) excitations coexisting at the same energy.  This is ensured
within the FSCM due to heavy particle charge conservation.  If there
are no true distinguishing quantum numbers of the excited state, decay
is possible and the singularity is rounded, as can be verified by
explicit calculation using, e.g. Fermi's golden rule.

We have already argued that for ideal conducting nanotubes and
symmetric semiconductor quantum wires, at least the first excited
subband is protected from decay in this way.  In any experiment,
various non-ideal perturbations will lead to some rounding.  Thermal
smearing (from thermal excitation both internally and in the tunneling 
lead) limits the resolution to $\epsilon \gtrsim k_B T$.  More
significant rounding arises from hybridization of the one-dimensional
electronic states with the bulk.  This is probably the most
significant effect in recent tunneling experiments using nanotubes on
metallic gold substrates.  This effect could be greatly reduced by
using an insulating substrate or better, a freely suspended tube.
Even for nanotubes on an insulating substrate, the asymmetry in
dielectric constants allows some degree of mixing of different subband states.
Fortunately, this is likely to be a weak effect due to the
delocalization of the subband states around the circumference of the
cylinder.  Finally, impurities or defects lead to smearing of the
ideal density of states on the scale of the inverse elastic scattering
time.

Additional processes left out of the FSCM may lead to significant
modifications of the physics, although they cannot remove the edge
singularity.  The forward scattering assumption itself is always
correct at energies sufficiently close to $\Delta$, since the heavy
particle cannot accomodate a large change in momentum without a
corresponding large increase in its energy.  There are, however,
orward scattering interactions outside the total charge channel.  The
most natural example is the Kondo interaction, ${\cal
  H}_K = - \lambda
(\psi^\dagger_R\vec{\sigma}\psi^{\vphantom\dagger}_R +
\psi^\dagger_L\vec{\sigma}\psi^{\vphantom\dagger}_L)\cdot d^\dagger
\vec{\sigma} d^{\vphantom\dagger}$.  For $m=\infty$, one has negligible effects for ferromagnetic
($\lambda>0$) coupling and perfect screening of the heavy spin for
antiferromagnetic coupling.  For finite $m$, one can
perform a perturbative renormalization group
calculation\cite{unpublished}\ which demonstrates that small
ferromagnetic coupling remains irrelevant for $m<\infty$.  Likewise,
for antiferromagnetic exchange, the problem scales to strong coupling.
The nature of the strong-coupling excitation(s) is not completely
clear, though preliminary investigations of some lattice models
suggest the formation of a propagating charge $e$ singlet particle.
As for the infinite mass model,\cite{Affleck94}\ we expect a large
(universal?)  contribution to the orthogonality exponent $\beta$ in
this case.  In the cases of nanotubes and quantum wires, as for most
itinerant systems, the bare Kondo couplings are ferromagnetic, and we
thus expect insignificant modifications in the spin channel.  The
antiferromagnetic case could perhaps be relevant, however, for certain
excited states in ladder materials.  The extra degeneracies in the
unoccupied subbands in nanotubes might also lead to some effects of
this type for {\sl pseudospin}, but only very close to the band edge
as all corrections to the FSCM are small in this case.\cite{KBF97}\ 

We conclude by summarizing the experimental predictions for nanotubes
and symmetric quantum wires.  While the tunneling density of states
for insulating systems should show the bare $(\epsilon-\Delta)^{-1/2}$
van Hove singularities (up to the two-particle continuum),
conducting systems should display a reduction of the edge singularity
at the first subband to $(\epsilon-\Delta)^{1/2-\beta}$.  For a
$1.4nm$ diameter nanotube, we estimate $\beta \approx 0.3$.  The
singularities at all higher subbands in metallic quantum wires
should be rounded.  


\acknowledgements

I would like to thank Aris Moustakas, Steve Simon, and Xenophon Xotos
for useful discussions.

\end{document}